# Research on Shape Mapping of 3D Mesh Models based on Hidden Markov Random Field and EM Algorithm

WANG Yong[1]    WU Huai-yu[2]

[1] (*University of Chinese Academy of Sciences, Beijing  100049, China*)

[2] (*Institute of Automation, Chinese Academy of Sciences, Beijing, 100080, China*)

**Abstract**   How to establish the matching (or corresponding) between two different 3D shapes is a classical problem. This paper focused on the research on shape mapping of 3D mesh models, and proposed a shape mapping algorithm based on Hidden Markov Random Field and EM algorithm, as introducing a hidden state random variable associated with the adjacent blocks of shape matching when establishing HMRF. This algorithm provides a new theory and method to ensure the consistency of the edge data of adjacent blocks, and the experimental results show that the algorithm in this paper has a great improvement on the shape mapping of 3D mesh models.

## 1 Introduction

Digital geometry processing of 3D mesh models has broad application prospects in the fields of industrial design, virtual reality, game entertainment, Internet, digital museum, urban planning and so on[1]. However the surface of 3D mesh models is usually bent arbitrarily, lack of continuous parameters, and has complex characterized details, as is quite different from the regular 2D image data with the uniform sampling, the data of 3D mesh models cannot be dealt with the classical orthogonal analysis tools directly. To meet the need of wide applications, researchers have proposed some processing algorithms to deal with 3D mesh models, such as surface reconstruction, mesh simplification, mesh smoothing, parametric, re-meshing, surface compression, mesh deformation and animation and so on. But these algorithms can only meet some specific requirements, the analysis and processing of 3D mesh models is still a public problem in the field of computer vision and computer graphics[2].

In intelligent analysis of 3D mesh data, how to "understand" the global and local 3D shapes is an important challenge for the analysis of 3D models, as is usually lacking in the most present methods. For example, in figure1, if the digital geometry processing framework can have global perspective and understand the shape globally, it can eliminate the interference and influence of shape analysis brought by rotation, translation, initial placement, bending deformation, different sampling rate, and different parameterization methods. (a1'/a2', b1'/b2') are the transformed models of (a1/a2, b1/b2) by the geometry processing framework of global perspective. It can be seen that the transformed models are much similar in their poses and shapes. So the difficulty of establishing the automatic matching between two different 3D shapes has been greatly reduced.

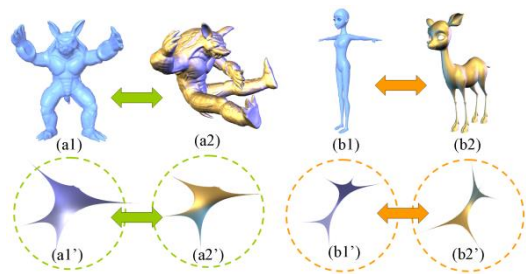

Figure 1 Matching between two different 3D shapes

How to establish the matching (or corresponding) between two different 3D shapes is a classical problem and has always been a hot issue [3, 4]. As is the prerequisite and basis for a large number of applications. These applications include: matching the shape template to multiple 3D data sets [5], shape blending [6], statistical shape analysis (such as principal component analysis), transfer texture and surface properties, surface classification and recognition, video tracking, facial animation based on facial expression, and so on. The shape matching can be realized either globally or locally. The former is

This work is supported by the joint research fund for UCAS and CAS institute (Y55201TY00).

computing and mapping the surface as a whole, while the latter is dividing the whole surface into blocks equally first of all, and then establishing the mapping of each block, finally putting the results of each block together to get the complete matching (also called mapping, or cross parameterization). The typical global methods include the iterative closest point method (ICP) [7] and its variants (e.g., [8,9], as are well-known global matching methods). However, the ICP algorithm is strongly dependent on the good initial shape, and is usually not suitable for the matching when the shapes are quite different. Therefore, many methods are relying on the user to manually place the identification points to guide the matching. For example, the work of [5] and [10], as are based on the template matching technology, establishing the shape matching (such as the human body model) according to the user's specified identification points . In addition to the direct global mapping, the global matching can also be indirectly established, which uses a temporary parameterized domain as a common domain [11,12]. For example, the shape of the disk topology can be mapped to a common plane, called the planar parameterization [11,13,14]. In addition, there is a method called spherical cross parameterization, in paper [12] a double mapping between the 0-genus closed surfaces is established based on it.

However, the biggest problems of shape matching only from a global point of view is that the geometric position of source / target curved face is usually different. The global solution does not take into account the particularity of each position, and the mapping (or more stringent bijective) got has a big twist. While the local algorithm can avoid this defect, as make little distortion. The local matching method usually first constructs an intermediate grid, and then computes the mapping between the intermediate grid and the source and the mapping between the intermediate grid and the target respectively. Finally, the two intermediate mappings are synthesized as the mapping between the source and the target. However, an important problem of local matching method is how to efficiently get the uniform block with high quality. Now there is no way to guarantee the optimal block, the usual method is to obtain the feasible sub block by some heuristic strategy. The efficiency and quality of the block algorithm is determined by the complexity of the heuristic strategy. Early methods such as shown in the paper [15] whose calculation speed is faster, but when dealing with complex surface their robustness is not enough. Because they use heuristic algorithms only in the cases that the source / target surface are shape similar. The methods in paper [6,11,16] use more sophisticated heuristic algorithms to ensure their robustness, but their time cost of computation is high. And when the number of blocks is large (>80), it takes a few hours of CPU time to get the legal blocks, as is intolerable in practical applications. More important, the local method also needs to ensure that the mapping of the neighboring blocks is continuous on their boundary. In paper [17] a method combining the global matching and partial matching is proposed, and a shape matching framework based on irregular middle domain is also presented, as constructs the discrete Laplace-Beltrami operator reflecting surface "local" parametric and local shape descriptors used for 3D model retrieval based on discrete exterior calculus method. This method converts the matching problem between two different (and complex) shapes to the matching problem between two similar (and simple) shapes. But how to ensure the edge data of adjacent blocks consistent is still difficult.

From the above analysis, we can see that if the global matching methods are used, when the difference between the two shapes is very large, the distortion error of the matching results is usually large. If the local matching methods are used, efficiently and robustly obtaining consistent chunking is a difficult problem to be solved. At present, most of the methods used are complex, time-consuming and highly skilled heuristic strategies. This is a challenge of real-time and robustness requirements for computer vision and virtual reality. This paper proposes a HMRF-EM algorithm based on Hidden Markov Random Filed model and EM algorithm, as introducing a hidden state random variable associated with the adjacent blocks of shape matching when establishing HMRF. This algorithm provides a new theory and method to ensure the consistency of the edge data of adjacent blocks. Figure 2 shows the framework of the whole research process.

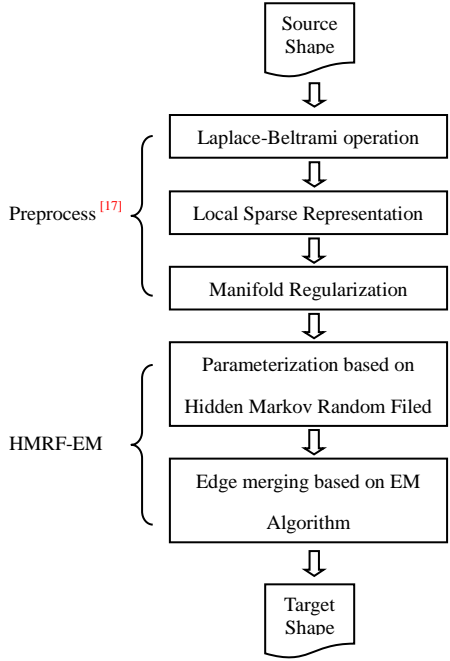

Figure 2 Architecture of research process

## 2 Hidden Markov Random Filed

The concept of Hidden Markov Random Field (HMRF) is derived from HMM[18], HMM is defined as a stochastic process that produces a Markov chain, and its state sequence can only be observed through observation sequence and cannot be obtained directly. Each observation sequence is assumed to be a state sequence of a random function. The underlying Markov chain changes its state according to an $l\times l$ transfer matrix, and Figure 3 shows its transfer process:

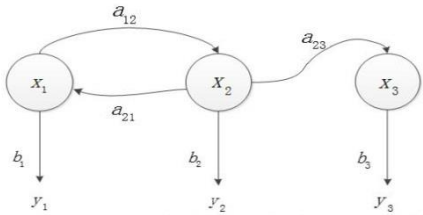

Figure 3 State diagram of Hidden Markov Random Field model

$x_i$ represents the hidden state, $y_i$ represents the observable output, $a_{ij}$ represents the transition probability, $b_i$ represents output probability, they all builds the Hidden Markov Random Field model。

The HMRF model contains a Hidden Markov Random Field and an Observable Random Field。The Hidden Random Field $X = \{x_i,\ i \in S\}$ is the assumption of finite state space $L$, and the state of $X$ is not observable. But the Observable Random Field $Y = \{y_i,\ i \in S\}$ is the random field of finite state space $D$, and each $y_i$ follows a known conditional probability distribution $p(y_i|x_i)$, as has the same function form $f(y_i; \theta_{x_i})$.

## 3 EM Algorithm

The Hidden Markov Random Field in HMRF model is used to describe the random process of the edge data of adjacent blocks, whose class labels are calculated based on the Observable Random Field. In this paper, the EM algorithm is used to realize the class label calculation. EM algorithm is an effective method to solve the optimization problem of latent variables. Given the Hidden Markov Random Field $X = \{x_i,\ i \in S\}$, the class label variable $Z$ is assumed to maximize the probability $p(X, Z)$. The equation of calculating the Maximum likelihood estimation of $p(X, Z)$ is as follows:

$$l(\theta) = \sum_{i \in S} \log p(x_i; \theta) = \sum_{i \in S} \log \sum_{z_i} p(x_i, z_i; \theta) \quad (1)$$

The first step to solve the equation is to take logarithm of the maximum likelihood, and the second step is to solve the joint probability distribution function of each possible Z of each sample. But it is generally very difficult to get the value of $\theta$ directly, because the variable Z is a latent variable. But if the variable Z is determined, the equation will be easily solved.

For each data $x_i$ belonging to variable Z, define $Q_i$ as the probability distribution function of Z (such as Gaussian distribution), so $Q_i$ meet the following equation:

$$\sum_{Z} Q_i(Z) = 1, \quad Q_i(Z) \geq 0 \quad (2)$$

Based on Jensen inequality, equation (1) can be transformed as follows:

$$\sum_{i \in S} \log p(x_i; \theta) = \sum_{i \in S} \log \sum_{z_i} p(x_i, z_i; \theta) \quad (3)$$

$$= \sum_{i \in S} \log \sum_{z} Q_i(z_i) \frac{p(x_i, z_i; \theta)}{Q_i(z_i)} \quad (4)$$

$$\geq \sum_{i \in S} \sum_{z} Q_i(z_i) \log \frac{p(x_i, z_i; \theta)}{Q_i(z_i)} \quad (5)$$

To solve the equation (3) based on EM algorithm, the first step is to initialize the parameter $\theta$, that is to initialize the probability distribution of Z. Then for each $i$, calculate $Q_i(z_i)$ based on the equation (6)

$$Q_i(z_i) = p(z_i|x_i; \theta) \quad (6)$$

The next step is to solve the equation (7)

$$\theta := \arg\max_{\theta} \sum_{i \in S} \sum_{z} Q_i(z_i) \log \frac{p(x_i, z_i; \theta)}{Q_i(z_i)} \quad (7)$$

This process is cycled, until the equation (5) is convergent.

## 4 HMRF-EM Algorithm

The advantage of EM algorithm is that it has reliable global convergence and faster convergence speed, but the maximum likelihood equation of hidden Markov model usually has multiple roots, which leads the EM algorithm into local extremum on the initial parameters setting. Therefore, we propose an EM algorithm based on hidden state random variables (Hidden Markov Random Field based EM Algorithm, HMRF-EM). The basic idea of the algorithm is to solve the parameter estimation problem based on incomplete data by recursive method. Specific steps of the algorithm are shown below:

---

**HMRF-EM Algorithm**

Input: coordination of the vector to be classified $Vec$, Class number $n$, the probability $p_{ij}$ of the $i$th vector belonging to the class $j$, covariance matrix of class $i$ $p_{cov}[i]$, mean vector of class $i$ $p_{mea}[i]$, the prior probability of class $i$ $p_{pior}[i]$.

output: mean and covariance of class $i$, the class has the biggest prior probability and the probability.

step:

1. initialization:
   1.1 $p_{ij}=0$;
   1.2 $p_{cov}[i]=1$;
   1.3 $p_{mea}[i]=i*2$;
   1.4 $p_{pior}[i]=1/n$;

2. For (the $i$th vector)
   2.1 if $i <$ the number of vectors to be classified
   2.2 then For (the $j$th class)
       2.2.1 if $j<n$
       2.2.2 then the vector probability density is calculated as
       $$p = \frac{|\sigma|^{\frac{-1}{2}} exp\left(-\frac{1}{2}(x-\mu_j)^T \sigma^{-1}(x-\mu_i)\right)}{\sqrt{2\pi}}$$
       2.2.3 $p_{ij}=p_{pior}[i]*p$;
       2.2.4 $Temp1+=p_{pior}[j]*p$;
   2.3 $E_{step}+=log(Temp1)$;

3. For (the $j$th class)
   3.1 if $j<n$
   3.2 then $Temp2+= p_{ij}$;
   3.3 For (the $i$th vector)
       3.3.1 if $i<$ the number of vectors to be classified
       3.3.2 then For (the $k$th vector)
           3.3.2.1 if $k<$ the dimension of vector
           3.3.2.2 then $p_{mea}[i][k]+= p_{ij}*Vec[i][k]$;
   3.4 $p_{pior}[i]= Temp2/Vec$;

4. return $p_{mea}[i]$, $p_{cov}[i]$, $p_{pior}[i]$ 和 $i$.

## 5 Conclusions

This paper focused on the research on shape mapping of 3D mesh models, and proposed a shape mapping algorithm based on Hidden Markov Random Field and EM algorithm, as introducing a hidden state random variable associated with the adjacent blocks of shape matching when establishing HMRF. Although the 3D geometric data intelligent analysis is widely used, but there are some disadvantages in processing the data of local part, so how to integrate a variety of methods to explore a variety of modal image segmentation will be the next research focus